\documentclass{aastex6}
\newcommand{\myfigA}{\ref{myfigA}}
\newcommand{\myfigB}{\ref{myfigB}}
\newcommand{\myfigC}{\ref{myfigC}}
\newcommand{\myfigD}{\ref{myfigD}}
\newcommand{\myfigE}{\ref{myfigE}}
\newcommand{\myfigF}{\ref{myfigF}}
\newcommand{\myfigG}{\ref{myfigG}}
\begin{document}
\title{First Electromagnetic Pulse Associated with a Gravitational-Wave Event: Profile, Duration, and Delay}
\author{Da-Bin Lin\altaffilmark{1}, Tong Liu\altaffilmark{2}, Jie Lin\altaffilmark{1}, Xiang-Gao Wang\altaffilmark{1}, Wei-Min Gu\altaffilmark{2}, and En-Wei Liang\altaffilmark{1}}
\altaffiltext{1}{Laboratory for Relativistic Astrophysics, Department of Physics, Guangxi University, Nanning 530004, China; lindabin@gxu.edu.cn}
\altaffiltext{2}{Department of Astronomy, Xiamen University, Xiamen, Fujian 361005, China}
\begin{abstract}
We study the first electromagnetic pulse after the gravitational wave chirp signal,
focusing on the profile and duration.
It is found that the light curve, especially the steep decay (SD) phase,
can be very different by adopting different viewing angle $\theta_{\rm view}$ on the jet shell.
For an on-axis jet with a power-law radiation spectrum,
the observed flux in the SD is proportional to $t_{\rm{obs}}^{-2-\beta}$
with $\beta$ being the spectral index and $t_{\rm{obs}}$ being the observer time.
Here, $t_{\rm{obs}}=0$ is set at the observed time of the jet ejected from the central engine.
The SD may become steep by increasing $\theta_{\rm view}$.
We also study the bolometric luminosity $L$ from a jet shell with a non-power-law radiation spectrum.
For an on-axis jet, $L{\propto}t_{\rm{obs}}^{-3}$ is found in the SD.
However, the SD is steeper than $L{\propto}t_{\rm{obs}}^{-3}$ for the radiation from an off-axis jet.
The higher value of $\theta_{\rm view}$ is, the steeper of SD would be.
Then, we suggest that the SD phase can be used to discriminate
an off-axis jet from an on-axis jet.
The reason for above behaviors is discussed.
In addition, we find that the duration of first electromagnetic pulse is
close to its peak time,
especially for $\theta_{\rm{view}}\sim20^\circ$.
This result is consistent with that found in GW~170817/GRB~170817A.
Thus, the jet corresponding to the prompt emission of GRB~170817A should be immediately ejected after the merger.
Our results also reveal that the duration of the first electromagnetic pulse can provide the information of the time to search gravitational waves.
\end{abstract}
%%%%%%%%%%%%%%%%%%%%%%%%%%%%%%%%%%%%%%%%%%%%%%%%%%%%%%%%%%%%%%%%%%%%%%%%%%%%%%%%%%%
%%%%%%%%%%%%%%%%%%%%%%%%%%%%%%%%%%%%%%%%%%%%%%%%%%%%%%%%%%%%%%%%%%%%%%%%%%%%%%%%%%%
\section{Introduction} \label{sec:intro}
%%%%%%%%%%%%%%%%%%%%%%%%%%%%%%%%%%%%%%%%%%%%%%%%%%%%%%%%%%%%%%%%%%%%%%%%%%%%%%%%%%%
%%%%%%%%%%%%%%%%%%%%%%%%%%%%%%%%%%%%%%%%%%%%%%%%%%%%%%%%%%%%%%%%%%%%%%%%%%%%%%%%%%%
For the first time, gravitational wave (GW) and electromagnetic (EM) wave
from a single source have been observed.
On August 17, 2017 at 12:41:04 UTC,
the Advanced Laser Interferometer Gravitational-wave Observatory and
the Advanced Virgo gravitational-wave detectors
have made their first observation on a binary neutron star (NS) merger (\citealp{Abbott_BP-2017a-Abbott_R,Abbott_BP-2017b-Abbott_R,Abbott_BP-2017c-Abbott_R,Abbott_BP-2017d-Abbott_R}).
The associated gravitational radiation event is known as GW~170817.
About 2 seconds after GW~170817,
the Fermi Gamma-ray Burst Monitor (GBM) has autonomously
detected a short gamma-ray burst (GRB), GRB~170817A,
from a location coincident with GW~170817 (\citealp{Goldstein_AE-2017}).
GRB~170817A was also detected by the International Gamma-Ray Astrophysics Laboratory (\citealp{Savchenko_V-2016-Ferrigno_C,He_XB-2017-Tam_PHT}).
These observations were followed by a detection of an optical counterpart,
SSS17a (now with the IAU identification of AT2017gfo, \citealp{Coulter_DA-2017}),
associated with the accompanying Macronova/kilonova
powered by the radioactive decay
of heavy elements formed in the NS-NS merger (\citealp{Li_LX-1998-Paczynski_B, Metzger_BD-2012-Berger_E, Berger_E-2013-Fong_W,Fernandez_R-2016-Metzger_BD,Ma_SB-2017-Lei_WH,Song_CY-2017-Liu_T,Liu_T-2017-Gu_WM}).
In addition, the accompanying Macronova/kilonova
was independently confirmed by several teams
(e.g., \citealp{Abbott_BP-2017a-Abbott_R,Arcavi_I-2017-Hosseinzadeh_G,
Hu_L-2017-Wu_X,Lipunov_VM-2017-Gorbovskoy_E,
Soares-Santos_M-2017-Holz_DE,
Tanvir_NR-2017-Levan_AJ,
Valenti_S-2017-David_Sand_J,Smartt_SJ-2017-Chen_TW,Troja_E-2017-Piro_L,Hallinan_G-2017-Corsi_A}).
The joint GW-EM detection has provided the first compelling observational evidence
on the relation of short GRBs and NS-NS mergers and
has led to the new era of gravitational-wave multi-messenger astrophysics.

The NS-NS merger is due to the loss of orbital
energy and angular momentum via gravitational radiation.
The merger of two NSs
can have four possible outcomes (\citealp{Abbott_BP-2017e-Abbott_R}):
(i) The prompt formation of a black hole (BH);
(ii) the formation of a hypermassive NS collapsing to a BH with $\lesssim$1~s;
(iii) the formation of a supramassive NS collapsing to a BH with $\sim 10-10^4$~s,
or
(iv) the formation of a stable NS.
Accretion onto the formed BH can launch a relativistic jet
and thus powers a GRB.
GRB~170817A was indeed found after GW~170817.
GRB~170817A triggered Fermi GBM with a duration of $T_{90}=2$~s (\citealp{Abbott_BP-2017a-Abbott_R,Fraija_N-2017b-Veres_P,He_XB-2017-Tam_PHT}).
The prompt $\gamma$-ray of GRB~170817A is faint
and peaks at $\sim 2.1$~s after the GM chirp signal (\citealp{Abbott_BP-2017a-Abbott_R,Fraija_N-2017b-Veres_P,He_XB-2017-Tam_PHT}).
In addition, a fast decay appears in the light curve of GRB~170817A after the peak time.
The physical origin of the prompt emission in GRB~170817A is still under debate.
The prompt emission in GRB~170817A may be formed in the photosphere of the jet (\citealp{Meng_YZ-2018-Geng_JJ}),
the internal shocks (\citealp{Murguia-Berthier_A-2017-Ramirez-Ruiz_E}),
the internal-collision-induced magnetic reconnection and turbulence (\citealp{Meng_YZ-2018-Geng_JJ}; \citealp{Zhang_B-2011-Yan_H}), or the external-reverse shock (RS, \citealp{Fraija_N-2017b-Veres_P}).
Except the radiation spectrum, these models should explain why there is not variability associated with this burst.
The emission from the reverse shock is exhibited as a single peak and thus can easily explain the single peaked behavior of GRB~170817A (\citealp{Fraija_N-2017b-Veres_P}).
The internal shock model, the pulse width is Doppler contracted, while the pulse separation is not.
Then a sizable increase in the viewing angle would cause significant
overlap between pulses and, as a result, the variability would be washed out (\citealp{Salafia_OS-2016-Ghisellini_G}).
The onset and peak time of GRB~170817A also drew a significant attention.
For example,
the delay of EM signal relative to the GW chirp signal has been used to constrain
the remnants of the NS-NS merger (\citealp{Granot_J-2017-Guetta_D}),
the weak equivalence principle (\citealp{Wei_JJ-2017-Zhang_BB,Shoemaker_IM-2017-Murase_K,Wang_H-2017-Zhang_FW}),
and the velocity of GW (\citealp{Wang_H-2017-Zhang_FW}).
In this paper,
however,
we would like to point out that
the peak time ($t_p$) of the first EM pulse is close to the duration ($T_{90}$) of this pulse
if the jet is immediately launched after the GW chirp signal.
In GW~170817/GRB~170817A, $t_p\sim T_{90}$ is found.
Then, we would like to believe that the jet associated with GRB~170817A is immediately launched after the NS-NS merger.
This result should be considered in utilizing the EM wave-GW delay in constraining the associated physical processes.
We also study the light curves of the first EM pulse.
It is found that the steep decay (SD) can be used to discriminate an off-axis jet emission from an on-axis jet emission.

The paper is organized as follows.
Since we study the radiation of the jet shell via a Monte Carlo method,
the procedure for simulating jet emission is presented in Section~\ref{Sec:Numerical calculation Procedure}.
The light curves and other results are presented in Section~\ref{Sec:Results}.
The conclusions and discussions are presented in Section~\ref{Sec:Conclusion}.

%%%%%%%%%%%%%%%%%%%%%%%%%%%%%%%%%%%%%%%%%%%%%%%%%%%%%%%%%%%%%%%%%%%%%%%%%%%%%%%%%%%%%%%%%%%%%%%%%%%%%%%%%
%%%%%%%%%%%%%%%%%%%%%%%%%%%%%%%%%%%%%%%%%%%%%%%%%%%%%%%%%%%%%%%%%%%%%%%%%%%%%%%%%%%%%%%%%%%%%%%%%%%%%%%%%
\section{Procedure for Simulating Jet Emission}\label{Sec:Numerical calculation Procedure}
We focus on the radiation of a spherical thin jet shell radiating from $r_0$ to $r_e$,
where $r_0$ and $r_e$ are estimated with respect to the jet base ($r=0$).
This process can appear in the internal shocks (\citealp{Rees_MJ-1994-Meszaros_P}), the internal-collision-induced magnetic reconnection and turbulence (\citealp{Zhang_B-2011-Yan_H}; \citealp{Deng_W-2015-Li_H}),
or the RS shock (e.g., \citealp{Shao_L-2005-Dai_ZG,Kobayashi_S-2007-Zhang_B}; \citealp{Fraija_N-2015,Fraija_N-2016-Lee_W,Fraija_N-2016-Lee_WH,Fraija_N-2017a-Veres_P,Fraija_N-2017b-Veres_P}),
especially for a radiating jet shell with a fast decaying behavior after the peak.
The jet shell (yellow region) is schematically shown in Figure~{\myfigA},
where the spherical coordinate with $r=0$ locating at the central engine of GRB
and $\theta=0$ being along the line of sight is adopted,
$\theta_{\rm jet}$ is the jet opening angle,
and
$\theta_{\rm view}$ is the viewing angle of the jet shell.
The jet shell is assumed uniform with sharp edges.

The radiation of our jet shell is computed via a Monte Carlo method
(e.g., \citealp{Lin_DB-2017a-Mu_HJ,Lin_DB-2017b-Mu_HJ}).
In brief, a number of emitters randomly distributed in the jet shell is used to simulate the radiation of the jet shell.
The radiation of an emitter in the jet shell comoving frame is assumed as (e.g., \citealp{Uhm_ZL-2015-Zhang_B})
\begin{equation}\label{eq:spec_power_single_ensemble}
P'(E') =P'_0\, H'(E'/E'_0),
\end{equation}
where $P_0^{\prime}$ describes the spectral power
and $E'_0$ is the characteristic photon energy of the radiation spectrum.
For the function form of $H'$, we study the following three cases:
\begin{equation}\label{Eq:H function}
\begin{array}{*{20}{c}}
{{\rm{Case\;(I)}}:}&{H'(x) = x^{\hat \beta+1},}\\
{{\rm{Case\;(II)}}:}&{H'(x) = \left\{ {\begin{array}{*{20}{c}}
{{x^{\hat \alpha + 1}}\exp ( { - x} ),}&{x \leqslant( {\hat \alpha- \hat \beta} ),}\\
{{{( {\hat \alpha - \hat \beta})}^{\hat \alpha - \hat \beta}}\exp ( {\hat \beta - \hat \alpha} ){x^{\hat \beta + 1}},}&{x \geqslant ( {\hat \alpha - \hat \beta}),}
\end{array}} \right.}\\
{{\rm{Case\;(III)}}:}&{H'(x) = x^{\hat \alpha}\exp (-x),}
\end{array}
\end{equation}
where $\hat \alpha$ and $\hat \beta$ are spectral indexes.
A photon in the comoving frame with energy $E'$ is boosted to $E=DE'/(1+z)$ in the observer's frame.
Here, $z$ is the redshift of the GRB and $D$ is the Doppler factor described as
\begin{eqnarray}\label{}
D ={\left[ {{\Gamma }(1 - {\upsilon _{\rm jet}}\cos \theta/c)} \right]^{ - 1}}
\end{eqnarray}
with $c$, $\theta$, $\upsilon_{\rm jet}=dr/dt$, and $\Gamma=1/\sqrt{1-(\upsilon_{\rm jet}/c)^2}$
being the light velocity, the latitude of the radiating emitter,
the jet shell velocity, and the jet shell Lorentz factor, respectively.
During the shell's expansion for $\delta t\; (\sim 0)$ from $r$ to $r+\upsilon_{\rm jet}\delta t$,
the observed spectral energy $\delta U$ from an emitter
into a solid angle $\delta \Omega$ is given as (\citealp{Uhm_ZL-2015-Zhang_B})
\begin{equation}\label{Eq:Numerical calculation-Flux}
\delta {U_E} (t_{\rm obs}) = \left({D^2}\delta \Omega\right)\left( \frac{{\delta t}}{\Gamma }\right)\frac{1}{{4\pi }}{P'_0}H'\left( {\frac{{E(1 + z)}}{{D{{E'}_0}}}} \right),
\end{equation}
where the radiation is assumed isotropically in the jet shell comoving frame (c.f. \citealp{Geng_JJ-2017-Huang_YF}).
The observed time of $\delta U$ is estimated with
\begin{equation}\label{Eq:t_obs}
t_{\rm obs} =\left\{ \int_{r_0}^{{r}} {[c - \upsilon_{\rm jet}]} \frac{{dl}}{c\upsilon_{\rm jet}(l) }+ \frac{r(1- \cos \theta)}{c}\right\}(1+z)+t_{\rm obs,r_0},
\end{equation}
where $\theta$ is the latitude of the radiating emitter.
In our work, $t_{\rm obs}=0$ is set at the observed time of the jet shell ejected
from $r=0$, and thus $t_{\rm obs,r_0}$ is the observed time of the emitter locating at $r_0$ and $\theta=0$.

The procedures of our simulations to obtain the observed flux is shown as follows.
Firstly, an expanding jet is modelled with a series of jet shells
at radius $r_0,\;r_1=r_0+\upsilon_{\rm jet}{\delta t},\;r_2=r_1+\upsilon_{\rm jet}{\delta t},\;\cdot\cdot\cdot,r_n=r_{n-1}+\upsilon_{\rm jet}{\delta t},\;\cdot\cdot\cdot$
appearing at the time $t=0{\rm s},\;{\delta t},\;{2\delta t},\;\cdot\cdot\cdot, n{\delta t},\;\cdot\cdot\cdot $, respectively.
During the shell's expansion for $\delta t$,
the shell moves from $r_{n-1}$ to $r_n$ with the same radiation behavior for emitters.
Secondly, we produce $N$ emitters centred at ($r_n$, $\theta_{\rm sh}$, $\varphi_{\rm sh}$) in spherical coordinates
with $\theta_{\rm sh}=0$ being along the axis of the jet shell,
where the value of $\cos\theta_{\rm sh}$ and $\varphi_{\rm sh}$ are randomly picked up from linear space of $[\cos\theta_{\rm jet},1]$ and $[0,2\pi]$, respectively.
Then, the value of $\cos\theta$ can be estimated with
\begin{equation}\label{}
\cos\theta  = \sin {\theta _{\rm sh}}\cos {\varphi _{{\rm sh}}}\sin {\theta _{{\rm{view}}}} + \cos {\theta _{{\rm sh}}}\cos {\theta _{{\rm{view}}}}.
\end{equation}
The observed spectral energy from an emitter
during the shell's expansion from $r_{n-1}$ to $r_n$ is calculated with Equation~(\ref{Eq:Numerical calculation-Flux}).
By discretizing the observer time $t_{\rm obs}$ into a series of short time intervals,
i.e., $[0, t_{{\rm obs},1}],\,[t_{{\rm obs},1},t_{{\rm obs},2}]\, \cdot\cdot\cdot\,,[t_{{\rm obs},k-1}, t_{{\rm obs},k}],\cdot\cdot\cdot$,
we can find the total observed spectral energy
\begin{equation}
\left. U_E \right|_{\left[t_{{\rm obs},k-1},t_{{\rm obs},k} \right)}=
\sum\limits_{t_{{\rm obs},k-1} \leqslant t_{\rm obs}<t_{{\rm obs},k}}{\delta {U_E} (t_{\rm obs})}
\end{equation}
in the time interval
$[t_{{\rm obs},k-1}, t_{{\rm obs},k}]$ based on Equations~ (\ref{Eq:Numerical calculation-Flux}) and (\ref{Eq:t_obs}).
Then, the observed flux $F_E$ at the observer time $(t_{{\rm obs},k-1}+ t_{{\rm obs},k})/2$ can be estimated as
\begin{equation}
F_E=\frac{{{{\left. U_E \right|}_{\left[t_{{\rm obs},k-1},t_{{\rm obs},k} \right)}}}}{{D_{\rm{L}}^2 (t_{{\rm obs},k}-t_{{\rm obs},k-1})\delta \Omega}},
\end{equation}
where $D_{\rm L}$ is the luminosity distance of the jet shell with respect to the observer.
The bolometric luminosity $L$ is written as
\begin{equation}
L=4\pi D_{\rm L}^2\int_0^\infty{F_EdE}.
\end{equation}

In our simulations,
the jet shell is assumed to radiate from $r_0=10^{12}\rm cm$ to $r_e=2r_0$
with a Lorentz factor $\Gamma=100$ and $E'_0=150{\rm keV}(1+z)/D_0$,
where $D_0=2\Gamma$ and $D_0=1/\Gamma[1-\beta\cos(\theta_{\rm view}-\theta_{\rm jet})]$ are adopted for $\theta_{\rm view}\leqslant\theta_{\rm jet}$ and $\theta_{\rm view}>\theta_{\rm jet}$, respectively.
Thus, $t_{\rm obs, r_0}=r_0/2\Gamma^2c$ can be estimated.
The value of $P'_0$ is assumed to increase with time $t'$ in the jet comoving frame,
i.e., $P'_0=P'_{0,0}t'$ with constant $P'_{0,0}$, and $t'=0$ is set at the situation that the jet shell arrives at $r_0$.
The value of $N>>1$, $\theta_{\rm jet}=5^\circ$, and $z=1$ are adopted and remain constant in our simulations.

%%%%%%%%%%%%%%%%%%%%%%%%%%%%%%%%%%%%%%%%%%%%%%%%%%%%%%%%%%%%%%%%%%%%%%%%%%%%%%%%%%%%%%%%%%%%%%%%%%%%%%%%%
%%%%%%%%%%%%%%%%%%%%%%%%%%%%%%%%%%%%%%%%%%%%%%%%%%%%%%%%%%%%%%%%%%%%%%%%%%%%%%%%%%%%%%%%%%%%%%%%%%%%%%%%%
\section{First Electromagnetic Pulse after Gravitational Chirp Signal}\label{Sec:Results}
In Figure~{\myfigB}, we show the light curves of the first EM pulse after
GW chirp signal by varying $\theta_{\rm view}$,
where Case (I) with $\hat \beta=-1.3$ is adopted in our simulations
and $t_p$ ($F_p$) is the peak time (flux) of the light curve.
The situations with $\theta_{\rm view}=0^\circ$, $5^\circ$, $10^\circ$, $20^\circ$, $40^\circ$, and $60^\circ$
are represented with the black, red, orange, yellow, green, and blue lines, respectively.
Similar to the results in \cite{Yamazaki_R-2003-Ioka_K},
the light curves from our simulations with different $\theta_{\rm view}$
can be very different, especially for those in the SD phase (i.e., $t_{\rm obs}>t_p$).
Then, we focus our attention on the SD.
For the light curves from our simulations, the SD is shaped by the shell curvature effect,
which is a combination of the time delay and the Doppler shifting of the intrinsic radiation spectrum for high
latitude ($\theta$) emission with respect to the emission from low latitude.
For the radiation from an on-axis jet shell with Case (I),
the light curves of the SD phase have been well studied (e.g., \citealp{Kumar_P-2000-Panaitescu_A}; \citealp{Dermer_CD-2004}; \citealp{Dyks_J-2005-Zhang_B};\citealp{Uhm_ZL-2015-Zhang_B};\citealp{Lin_DB-2017a-Mu_HJ,Lin_DB-2017b-Mu_HJ}).
The well-known relation, $\alpha=\beta+2$, has been derived in the SD phase,
where the value of $-\alpha$ is the power law decay index (i.e., $F\propto t_{\rm obs}^{-\alpha}$)
and $\beta$ is the spectral index.
Then, we plot the light curve (cyan dashed line) of $F\propto t_{\rm obs}^{-\beta-2}$
in the right panel of Figure~{\myfigB},
where $\beta=-\hat \beta$ is obtained for the simulations with Case (I).
It can be found that
the light curve in the SD phase may deviate from
$F\propto t_{\rm obs}^{-\beta-2}$ for the simulations with high $\theta_{\rm view}$.
The higher value of $\theta_{\rm view}$ we adopt, the more obvious the deviation would appear.
Then, we suggest that the SD can be used to
discriminate an off-axis jet from an on-axis jet.
We also study the light curves in the situations with Case (II) or (III).
In these situations, however, the relation of $\alpha$ and $\beta$ is complex in the SD phase
due to the evolution of $\beta$ (see \citealp{Lin_DB-2017a-Mu_HJ,Lin_DB-2017b-Mu_HJ}).
Then, we study the light curves of the bolometric luminosity $L$ in Figure~{\myfigC},
where $L_p$ is the peak luminosity.
For a SD shaped by the curvature effect,
the light curves can be described as $L\propto t_{\rm obs}^{-3}$ for an on-axis jet shell.
Then, we show the relation of $L\propto t_{\rm obs}^{-3}$ in Figure~{\myfigC}
with cyan dashed lines.
The meanings of other lines in Figure~{\myfigC} are the same as those in Figure~{\myfigB}.
One can find that the luminosity in the SD phase
also deviates from $L\propto t_{\rm obs}^{-3}$ for the situations with high $\theta_{\rm view}$.
The higher value of $\theta_{\rm view}$ we take, the more obvious the deviation would be.
We also perform spectral fittings and obtain the value of $E_0$ ($E_c$) for the simulations with Case (II) or (III).
The obtained $E_0$ ($E_c$) is shown in the left panels of Figure~{\myfigD}
and almost consistent with $E_0\propto t_{\rm obs}$ ($E_c \propto t_{\rm obs}$) in the SD phase,
where $E_0\propto t_{\rm obs}$ ($E_c \propto t_{\rm obs}$) is shown with a cyan dashed line.
The right panels of Figure~{\myfigD} show the relations of $L-E_c$ (upper-right panel) and $L-E_0$ (lower-right panel) in the SD phase.
One can find that the relation of $L-E_0$ ($L-E_c$) deviates from $L\propto E_0^3$ ($L\propto E_c^3$)
for the simulations with high $\theta_{\rm view}$,
where $L\propto E_0^3$ ($L\propto E_c^3$) is shown with a cyan dashed line.
In Figure~{\myfigD}, the meanings of other lines are the same as those in Figure~{\myfigB}.
The higher value of $\theta_{\rm view}$ is, the steeper of $L-E_c$ ($L-E_0$) would be.
Then, we conclude that the SD can be used to estimate the viewing angle $\theta_{\rm view}$.

The reason for above found deviations is shown as follows.
According to Figures~{\myfigB} and {\myfigC},
one can find that
the deviation of the SD from the cyan dashed line
only become evident in the simulations with $\theta_{\rm view}>\theta_{\rm jet}$.
Then, we discuss the situations with $\theta_{\rm view}>\theta_{\rm jet}$.
For the radiation from the dotted region (see Figure~{\myfigA}),
the evolution of $L$ can be described as (\citealp{Lin_DB-2017a-Mu_HJ})
\begin{equation}\label{Eq:Luminosity}
L \propto \left(1 + \frac{t_{\rm obs}-t_0}{t_c} \right)^{ - 3},
\end{equation}
where
\begin{equation}\label{Eq:t_c}
t_c (r)=\left\{ {\frac{r}{{2{\Gamma ^2}c}} + \frac{r}{c}\left[ {1 - \cos \left( {{\theta _{{\rm{view}}}} - {\theta _{{\mathop{\rm jet}\nolimits} }}} \right)} \right]} \right\}\left( {1 + z} \right)
\end{equation}
is the angular spreading timescale for the radiation from the dotted region and
$t_0(=t_c)$ is the observed time of the first photon from the dotted region.
The dotted region in Figure~{\myfigA} is an annulus
with $\theta\in [\theta _{\rm view}- \theta _{\rm jet},\theta _{\rm view}+\theta _{\rm jet}]$
and swaddled by two longitude $l_{\rm E}$ and $l_{\rm W}$,
where $l_{\rm E}$ and $l_{\rm W}$ are tangent to the jet shell at the latitude $\theta_{\rm c}$.
From Figure~{\myfigA}, one can find that
the covered dotted region by the jet shell increases with $\theta$ for $\theta<\theta_c$
and decreases with $\theta$ for $\theta>\theta_c$.
Then, the evolution of $L$ relative to $t_{\rm obs}$
may be shallower than that of Equation~(\ref{Eq:Luminosity}) at the early phase of the SD
and steeper than that of Equation~(\ref{Eq:Luminosity}) at the later phase of the SD.
These behaviors can be found in the situations with $\theta_{\rm view}=10^\circ$ and $20^\circ$.
Furthermore, the total duration of the SD phase can be read as
\begin{equation}\label{Eq:t_duration}
t_d=(1+z)\frac{r}{c}[\cos(\theta_{\rm view}-\theta_{\rm jet})-\cos(\theta_{\rm view}+\theta_{\rm jet})].
\end{equation}
A sharp cutoff would appear in the light curve at $t_{\rm obs}\gtrsim t_d$.
It should be noted that the value of $t_d/t_c$ decreases with increasing $\theta_{\rm view}$
based on Equations~(\ref{Eq:t_c}) and (\ref{Eq:t_duration}).
This behavior can be found in Figure~{\myfigE},
where the black, red, and blue solid lines represent the value of $t_d/t_c$
calculated with $\Gamma=50$, 150, 450, respectively.
In addition, the $\theta_{\rm jet}=5^\circ$ ($10^\circ$) is adopted in the left (right) panel
and the green dashed line represents $t_d=t_c$.
According to this figure,
$t_d<t_c$ appears at $\theta_{\rm view}\gtrsim 30^\circ$ ($55^\circ$) in the simulations with $\theta_{\rm jet}=5^\circ$ ($10^\circ$).
Then, the SD would quickly enter into the sharp cutoff phase for
$\theta_{\rm view}\gtrsim 30^\circ$ ($55^\circ$) in the simulations with $\theta_{\rm jet}=5^\circ$ ($10^\circ$).
This behavior can be easily found in the light curves
from the simulations with $\theta_{\rm view}=40^\circ$ or $60^\circ$.
By varying $\theta_{\rm jet}$, we show the value of $\theta_{\rm view}$ satisfying $t_d=t_c$ in Figure~{\myfigF},
where $\Gamma = 150$ is adopted in our calculation.
One should note that the value of $\Gamma (\gtrsim 50)$ does not affect the obtained $\theta_{\rm view}$,
which can be found in Figure~{\myfigE}.
Figure~{\myfigF} suggests that the sharp cutoff is more likely to appear
in the SD shaped by an off-axis jet shell with a low $\theta_{\rm jet}$.
One would note that the onset of afterglow may overlap in time with the SD of the prompt emission (\citealp{O'Brien_PT-2006-Willingale_R}; \citealp{Fraija_N-2017-Lee_WH}), especially for an on-axis jet.
For an off-axis jet shell, the variability timescale of the prompt emission, the peak time of the afterglow onset,
and the separation between the prompt emission and the afterglow onset is proportional to
$\Gamma[1-\upsilon_{\rm jet}\cos(\theta_{\rm view}-\theta_{\rm jet})/c]$.
As discussed above, however, the sharp cutoff may quickly appear in the situation with an off-axis jet and
thus the SD and the afterglow onset can be well separated.
That is to say, one can easily identify the SD of prompt emission and the afterglow onset for an off-axis jet.

0

In our simulations, the jet shell is radiating from $r_0$ to $r_e$ with $P'_0$ increasing with $t'$.
In this situation, the SD is dominated by the radiation from the jet shell located at $r_e$
and the peak time of luminosity would be at $\sim t_0(r_e)$, i.e., $t_p\sim t_0(r_e)=t_c(r_e)$.
This is to say, the value of the peak time for the first EM pulse is close to the decay timescale of the SD phase.
Then, one would expect that the duration of the first EM pulse would
be close to the value of the peak time of the first EM pulse.
In Figure~{\myfigG}, we study the relation of $T_{90}$ and $t_{\rm p}$
for the light curves plotted in Figurs~{\myfigB} and {\myfigC},
where the symbols of ``$\circ$'', ``$+$'', and ``$\times$'' represent
the results from the simulations with Case (I), (II), and (III), respectively.
From Figure~{\myfigG}, the $0.4\lesssim T_{90}/t_{p} \lesssim 3$ can be found.
In addition, $T_{90}\sim t_{p}$ can be easy found for $\theta_{\rm view}=20^\circ$.
This result is consistent with the EM wave-GW delay found in GW~170817/GRB~170817A.
The delay of the first EM pulse with respect to the GW chirp signal can be decomposed into three components:
(1) the time for the jet formation, which can be neglected for GW~170817/GRB~170817A (e.g., \citealp{Zhang_BB-2017-Zhang_B});
(2) the time for the jet propagating from the central engine to the dissipation location;
(3) the geometrical delay, $\Delta t_{\rm g}=(1 + z)[1 - \cos(\theta _{\rm view} - \theta _{\rm jet})]r/c$,
which is induced by the additional path of the jet edge relative to the light of sight
and only appears in the situation of $\theta_{\rm view}>\theta _{\rm jet}$.
For a high viewing angle $\theta_{\rm view}$, the delay induced by (2) is less than that induced by (3).
Then, the delay of the first EM pulse relative to the GM chirp signal would be dominated by $\Delta t_{\rm g}\sim t_c$.
This is the reason for the consistency of our obtained delay with that found in GW~170817/GRB~170817A.
Then, we conclude that the lead of GW chirp signal with respect to the EM pulse is around the duration of the first EM pulse.

%%%%%%%%%%%%%%%%%%%%%%%%%%%%%%%%%%%%%%%%%%%%%%%%%%%%%%%%%%%%%%%%%%%%%%%%%%%%%%%%%%%%%%%%%%%%%%%%%%%%%%%%%
%%%%%%%%%%%%%%%%%%%%%%%%%%%%%%%%%%%%%%%%%%%%%%%%%%%%%%%%%%%%%%%%%%%%%%%%%%%%%%%%%%%%%%%%%%%%%%%%%%%%%%%%%
\section{Conclusions and Discussions}\label{Sec:Conclusion}
In this work, we study the light curve of the first electromagnetic pulse after gravitational wave chirp signal.
We find that the light curve,
especially the steep decay phase,
can be very different for different viewing angle $\theta_{\rm view}$.
In our work, the SD is shaped by the shell curvature effect.
For the radiation from an on-axis jet shell with a power-law intrinsic radiation spectrum,
the light curves in the SD well follow $F_\nu\propto t_{\rm obs}^{-2-\beta}$ with $\beta$ being the spectral index.
However, the light curves deviate from $F_\nu\propto t_{\rm obs}^{-2-\beta}$ for the situation with high $\theta_{\rm view}$.
The higher value of $\theta_{\rm view}$ is, the more obvious of the deviation would be.
We also study the bolometric luminosity from an radiating jet shell with a non-power-law radiation spectrum.
In this situation, the bolometric luminosity in the SD can be described as $L\propto t_{\rm obs}^{-3}$
for an on-axis jet shell.
For an off-axis jet shell, the luminosity in the SD also deviates from $L\propto t_{\rm obs}^{-3}$.
The higher value of $\theta_{\rm view}$ we adopt, the more obvious the deviation would appear.
Then, we conclude that the SD can be used to discriminate between an on-axis jet radiation and an off-axis jet radiation.
We also present the explanation for the above found deviations.
In addition, it is found that the duration of the first EM pulse is close to the value of its peak time,
especially for the situation with $\theta_{\rm view}\sim20^\circ$.
This result is consistent with the EM wave-GW delay found in GW~170817/GRB~170817A.
Thus, the jet corresponding to the prompt emission of GRB~170817A should be immediately ejected after the NS-NS merger.
Our results also reveal that the duration of the first EM pulse can provide the searching time for gravitational chirp signal
in a GW event associated with a GRB.

A structured jet is usually used to fit the afterglow of GW~170817/GRB~170817A (e.g., \citealp{Lazzati_D-2017-Perna_R,Meng_YZ-2018-Geng_JJ,Lyman_JD-2018-Lamb_GP}).
{\bf To explain the EM wave-GW delay found in GW~170817/GRB~170817A,
however, a significant delay between the NS-NS merger and the launch of the jet is required in this scenario
(e.g., \citealp{Meng_YZ-2018-Geng_JJ}).
In this situation, the value of $T_{90}$ would be significantly less than that of $t_p$ based on our obtained results.
This behavior is inconsistent with the observations (i.e., $T_{90}\sim t_p$).}
Furthermore, it should be noted that
the jet structure in the prompt emission phase and that in the later afterglow phase
can be very different due to the lateral expansion of the jet (e.g., \citealp{Rhoads_JE-1997,Rhoads_JE-1999,Huang_YF-2000-Gou_LJ}).
Then, a structured jet may well explain the afterglow emission
but is not a required ingredient in explaining the prompt emission.
Our found relation between $T_{90}$ and $t_p$ is applicable
for a radiating jet shell in the internal shocks (\citealp{Rees_MJ-1994-Meszaros_P}),
the internal-collision-induced magnetic reconnection and turbulence (\citealp{Zhang_B-2011-Yan_H}; \citealp{Deng_W-2015-Li_H}), or the RS shock (e.g., \citealp{Shao_L-2005-Dai_ZG,Kobayashi_S-2007-Zhang_B,Fraija_N-2015,Fraija_N-2016-Lee_W,Fraija_N-2016-Lee_WH,Fraija_N-2017a-Veres_P,Fraija_N-2017b-Veres_P}),
especially for those with a fast decaying behavior after the peak.
The light curves formed in the external-forward shock always
has a normal decay with power-law decay index $\sim -1.2$ (e.g., \citealp{Zhang_B-2006-Fan_YZ}).
In this situation, one could not define $T_{90}$ and the value of $T_{90}/t_p$.
Then, our model is not applicable for the external-forward shock.
Several GRBs with a flash formed in the RS shock are identified,
e.g., GRBs
990123 (\citealp{Akerlof_C-1999-Balsano_R}),
041219A (\citealp{Blake_CH-2005-Bloom_JS};\citealp{Vestrand_WT-2005-Wozniak_PR}),
050820A (\citealp{Vestrand_WT-2006-Wren_JA}),
090102 (\citealp{Steele_IA-2009-Mundell_CG}),
090510 (\citealp{Fraija_N-2016-Lee_WH,Ackermann_M-2010-Asano_K}),
110731A (\citealp{Fraija_N-2015,Ackermann_M-2013-Ajello_M}),
130427A (\citealp{Vestrand_WT-2014-Wren_JA}; \citealp{Fraija_N-2016-Lee_W}),
140512A (\citealp{Huang_XL-2016-Xin_LP}), and
160625B (\citealp{Zhang_BB-2016-Zhang_B,Lu_HJ-2017-Lu_J, Fraija_N-2017a-Veres_P}).
Table~\ref{MyTabA} shows the value of $T_{90}$, $t_p$, and $T_{90}/t_p$ for the RS flash in these bursts,
where the trigger time of the burst is set as the zero time.
The relation of $T_{90}\sim t_p$ can be easily found.
We would like to point out that, for flashes formed in a same RS,
one can have $T_{90}\sim 3t_p$ for the flash observed in the optical band if
$T_{90}\sim 0.5t_p$ is obtained for the flash observed in the Fermi-LAT band.
This conclusion is obtained based on the observations of GRB~130427A.
In this burst, the $T_{90}\sim 4.66t_p$ ($T_{90}\sim 0.26t_p$) is found for the optical (Fermi-LAT) flash,
which is explained through synchrotron (synchrotron self-Compton) emission from the RS.
Then, one can have $T_{90}\sim 3t_p$ for the RSs listed in Table~\ref{MyTabA} and observed in the optical band.
These results reveal that the jet in GRBs listed in Table~\ref{MyTabA} would be in the on-axis situation rather than off-axis situation.

The relation between $T_{90}$ and $t_p$ can be used to estimate the ejection time of the jet.
In Table~\ref{MyTabA}, GRBs~050820A and 160625B are not discussed
since the trigger time of the burst is not the zero time of our focused RS flash.
The energy injection into the external shock is discussed in these two bursts, e.g.,
\cite{Vestrand_WT-2006-Wren_JA}, \cite{Fraija_N-2017a-Veres_P}, and \cite{Lin_DB-2018-Huang_BQ}.
Central engines of GRBs may be intermittent and launch several episodes of ejecta separated
by a long quiescent interval. In this scenario, an external shock is formed due to the propagation of the first
launched ejecta into the circum-burst medium and the later launched ejecta may interact with the external shock at
a later period. Then, the onset of the dominant afterglow component should be referenced to the time of the onset of the dominant $\gamma$-ray pulse, especially for GRBs with a weak precursor, e.g., GRB~160625B.
Based on the observations of the optical flash (\citealp{Zhang_BB-2016-Zhang_B}; \citealp{Lu_HJ-2017-Lu_J}), we have $T_{90}\sim 73$~s and $t_p=207$~s.
Then, the ejected time of this ejecta is at around $t_p-T_{90}/3\sim 183$~s,
which is the beginning of the second sub-burst in GRB~160625B and consistent with that found in \cite{Fraija_N-2017a-Veres_P}.
It is interesting to point out that the ejection time of an on-axis jet shell
can be exactly estimated by fitting the SD (\citealp{Lin_DB-2017a-Mu_HJ}).
In \cite{Lin_DB-2017a-Mu_HJ},
we move the zero time $t_0$ to a certain time in the SD phase
and derive an analytical formula to describe the bolometric luminosity evolution,
i.e., Equation~(\ref{Eq:Luminosity}) in this paper.
For a coasting jet shell, the $t_0-t_c$ is corresponding to the observed time of the jet shell ejected from the central engine.
One can used Equation~(\ref{Eq:Luminosity}) to fit the SD of the first EM pulse and obtain the value of $t_c$.
If $t_0-t_c$ lags behind the observed time of GW chirp signal,
the first EM pulse may be produced in an off-axis jet shell.
This is another method to judge the jet being on-axis or off-axis.

%%%%%%%%%%%%%%%%%%%%%%%%%%%%%%%%%%%%%%%%%%%%%%%%%%%%%%%%%%%%%%%%%%%%%%%%%%%%%%%%%%%%%%%%%%%%%%%%%%%%%%%%%
%%%%%%%%%%%%%%%%%%%%%%%%%%%%%%%%%%%%%%%%%%%%%%%%%%%%%%0%%%%%%%%%%%%%%%%%%%%%%%%%%%%%%%%%%%%%%%%%%%%%%%%%%%
\acknowledgments
This work is supported by
the National Basic Research Program of China
(973 Program, grant No. 2014CB845800),
the National Natural Science Foundation of China (grant Nos. 11773007, 11403005, 11533003, 11673006, 11573023, 11473022),
the Guangxi Science Foundation (grant Nos. 2016GXNSFDA380027, 2016GXNSFFA380006),
the Special Funding for Guangxi Distinguished Professors (Bagui Yingcai \& Bagui Xuezhe),
the Innovation Team and Outstanding Scholar Program in Guangxi Colleges,
and
the One-Hundred-Talents Program of Guangxi colleges.

%%%%%%%%%%%%%%%%%%%%%%%%%%%%%%%%%%%%%%%%%%%%%%%%%%%%%%%%%%%%%%%%%%%%%%%%%%%%%%%%%%%%%%%%%%%%%%%%%%%%%%%%%%%%%%%%%%%%%%
%%%%%%%%%%%%%%%%%%%%%%%%%%%%%%%%%%%%%%%%%%%%%%%%%%%%%%%%%%%%%%%%%%%%%%%%%%%%%%%%%%%%%%%%%%%%%%%%%%%%%%%%%%%%%%%%%%%%%%
%%%%%%%%%%%%%%%%%%%%%%%%%%%%%%%%%%%%%%%%%%%%%%%%%%%%%%%%%%%%%%%%%%%%%%%%%%%%%%%%%%%%%%%%%%%%%%%%%%%%%%%%%%%%%%%%%%%%%%%%%%%%%%%%%%%%%%%%%%%%%%%%%%%%%%%%%%
%%%%%%%%%%%%%%%%%%%%%%%%%%%%%%%%%%%%%%%%%%%%%%%%%%%%%%%%%%%%%%%0%%%%%%%%%0%%%%%%%%0%%%%%%%%%%%%%%%%%%%%%%%%%%%%%%%%%%%%%%%%%%%%%%%%%%%%%%%%%%%%%%%%%%%%%%%%%%

\clearpage
\begin{figure}
\centering
\begin{tabular}{c}
\includegraphics[angle=0,scale=1.0,trim=00 0 00 0,clip]{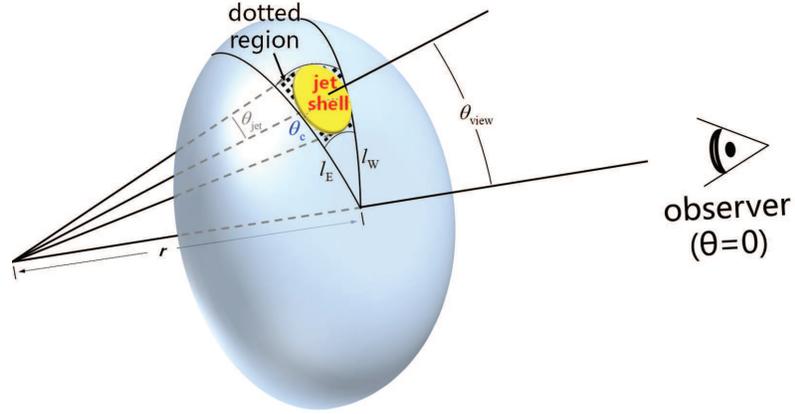}\\
\end{tabular}
\caption{Illustration of our radiating jet shell (yellow region).
Here, the spherical coordinate with $r=0$ locating at the central engine of GRB
and $\theta=0$ being along the line of sight is adopted,
$\theta_{\rm jet}$ is the jet opening angle,
and
$\theta_{\rm view}$ is the viewing angle of the jet shell axis.
The dotted region is an annulus
with $\theta\in [\theta _{\rm view}- \theta _{\rm jet},\theta _{\rm view}+\theta _{\rm jet}]$
and swaddled by two longitudes $l_{\rm E}$ and $l_{\rm W}$,
where $l_{\rm E}$ and $l_{\rm W}$ are tangent to the jet shell at the latitude $\theta_{\rm c}$.}\label{myfigA}
\end{figure}
%%%%%%%%%%%%%%%%%%%%%%%%%%%%%%%%%%%%%%%%%%%%%%%%%%%%%%%%%%%%0000000000000000000000000000000000000000000000000000000000000000000000000000000
%%%%%%%%%%%%%%%%%%%%%%%%%%%%%%%%%%%%%%%%%%%%%%%%%%%%%%%%%%%%000000000000000000000000000000000000000000000000000000000000000000000000

\clearpage
\begin{figure}
\centering
\begin{tabular}{cc}
\includegraphics[angle=0,scale=0.3,trim=00 0 00 0,clip]{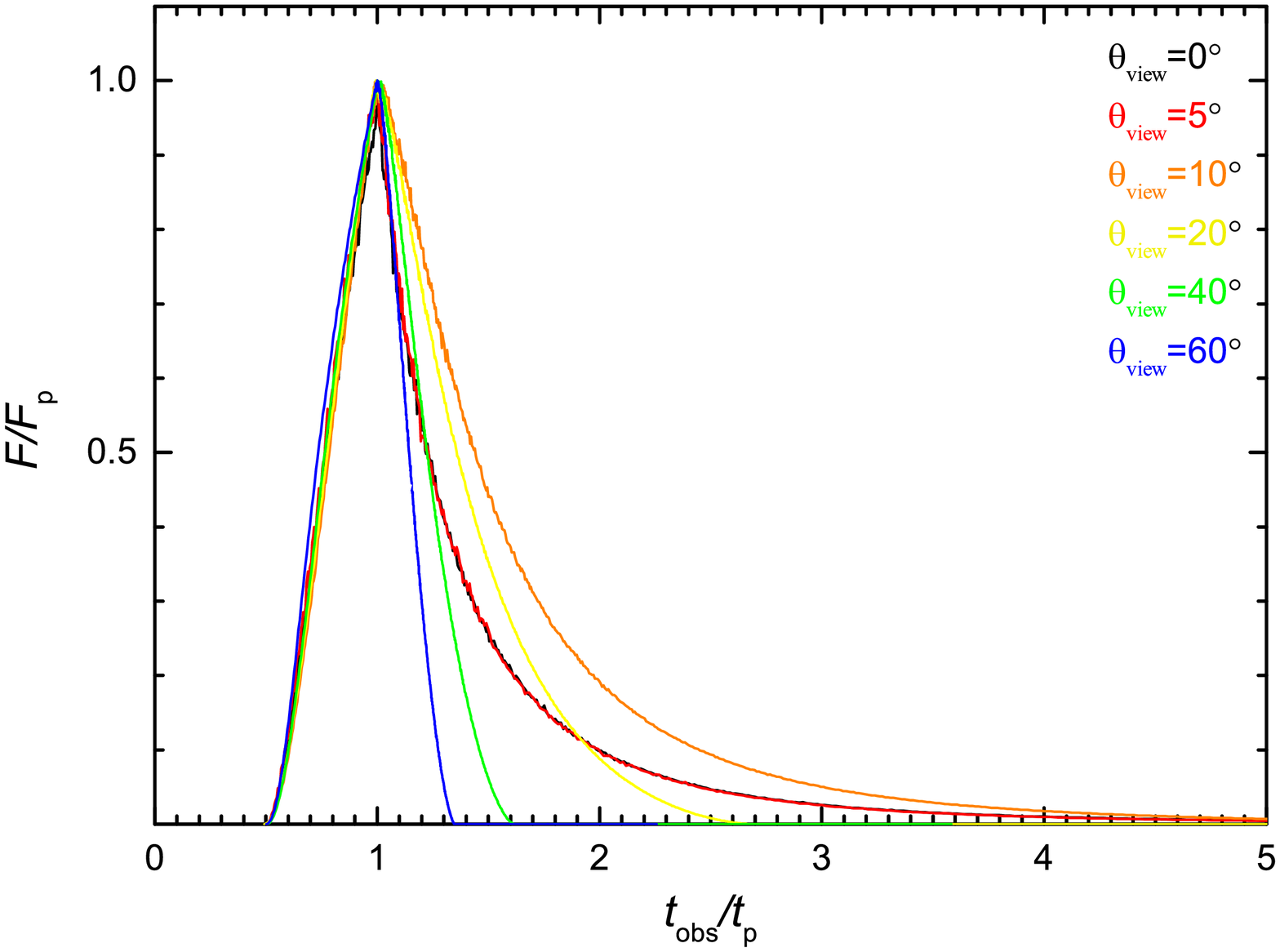} & \includegraphics[angle=0,scale=0.3,trim=00 0 00 0,clip]{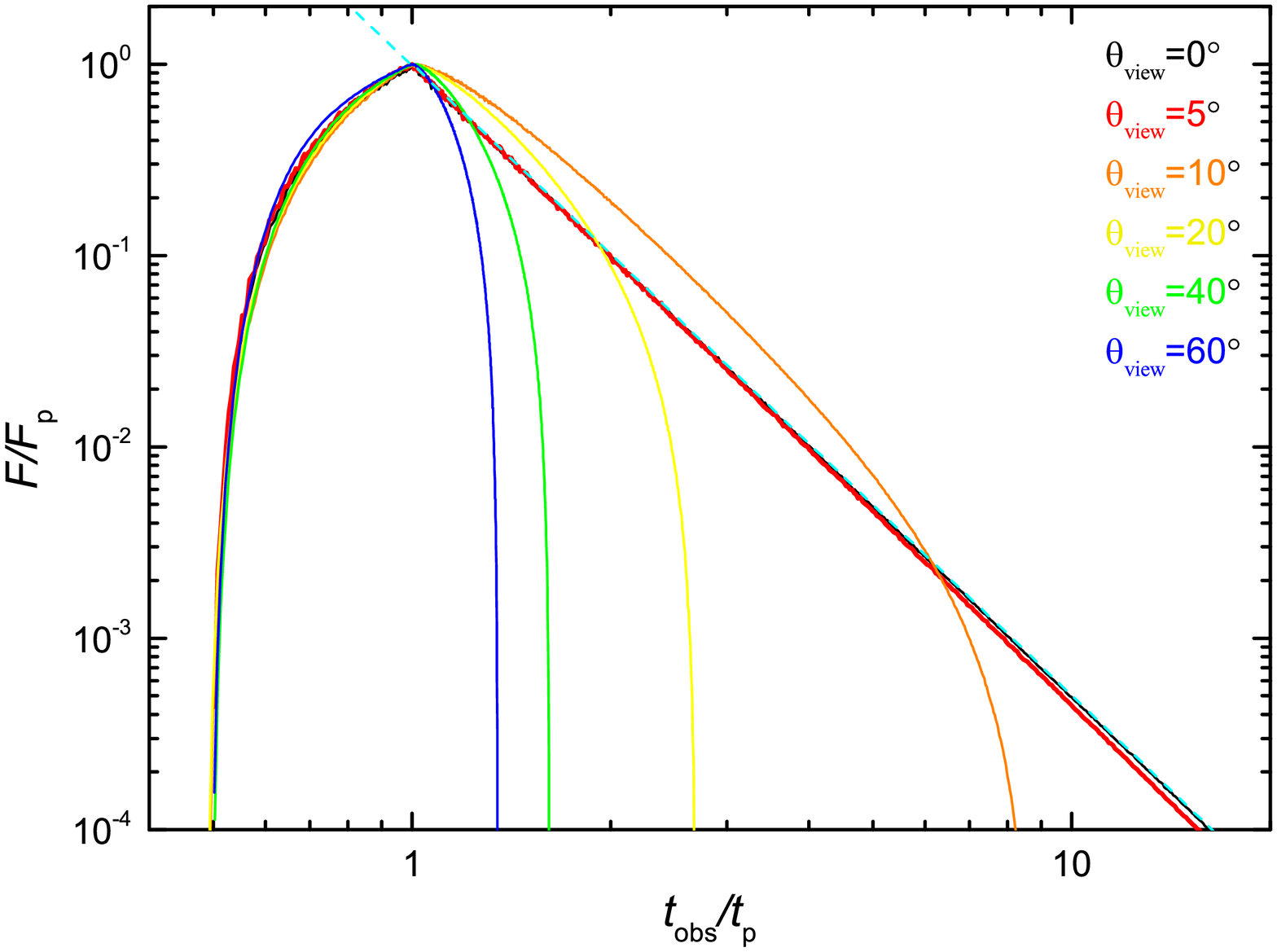}\\
\end{tabular}
\caption{Light curves of the first electromagnetic pulse, where the black, red, orange, yellow, green, and blue solid lines represent the situations with $\theta_{\rm view}=0^\circ,\, 5^\circ,\, 10^\circ,\, 20^\circ,\, 40^\circ$, and $60^\circ$, respectively. The cyan dashed line in the right panel plots the relation of $F\propto t_{\rm obs}^{-2-\beta}$.}\label{myfigB}
\end{figure}
%%%%%%%%%%%%%%%%%%%%%%%%%%%%0%%%%%%%%%%%%%%%%%%%%%%%%%%%%%%%%00000000000000000000000000000000000000000000000000000000000000000000000
%%%%%%%%00%%%%%%%%%%%%%%%%%%%%%%%%%%%%%%%%%000%%%%%%%%%%%%%%%%%%%00000000000000000000000000000000000000000000000000000000000000000000000

\clearpage
\begin{figure}
\centering
\begin{tabular}{cc}
\includegraphics[angle=0,scale=0.3,trim=00 0 00 0,clip]{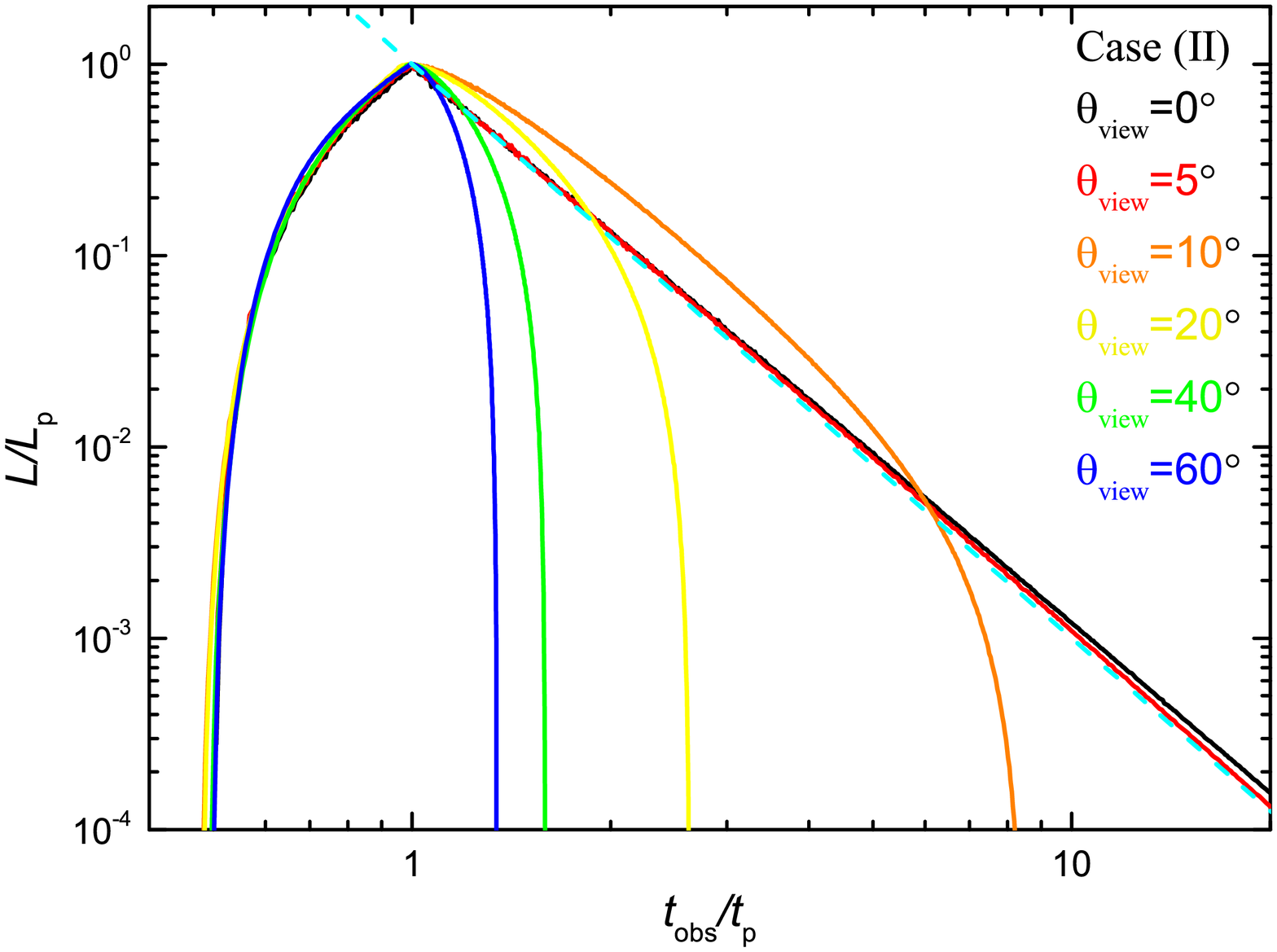} & \includegraphics[angle=0,scale=0.3,trim=00 0 00 0,clip]{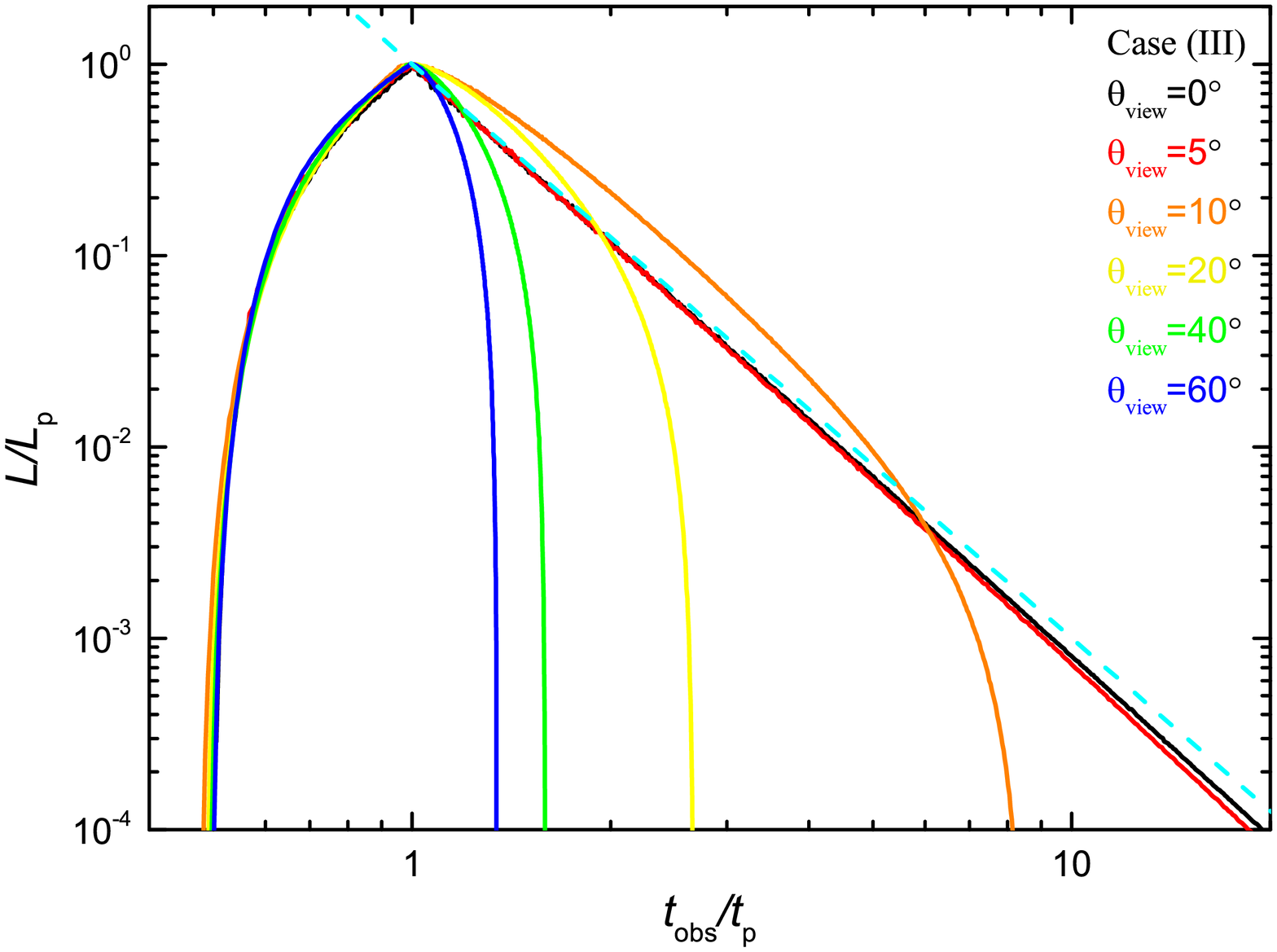}\\
\end{tabular}
\caption{Luminosity of an radiating jet shell with Case (II) (left panel) and Case (III) (right panel),
where the black, red, orange, yellow, green, and blue solid lines represent the situations with $\theta_{\rm view}=0^\circ,\, 5^\circ,\, 10^\circ,\, 20^\circ,\, 40^\circ$, and $60^\circ$, respectively.
The cyan dashed lines plot the relation of $L\propto t_{\rm obs}^{-3}$.}\label{myfigC}
\end{figure}
%%%%%%%%%%%%%%%%%%%%%%%%%%%%%%%%%%%%%%%%%%%%%%%%%%%%%%%%%%%%0000000000000000000000000000000000000000000000000000000000000000000
%%%%%%%%%%%%%%%%%%%%%%%%%%%%%%%%%%%%%%%%%%%%%%%%%%%%%%%%%%%%0000000000000000000000000000000000000000000000000000000000000000000000000000

\clearpage
\begin{figure}
\centering
\begin{tabular}{cc}
\includegraphics[angle=0,scale=0.3,trim=00 0 00 0,clip]{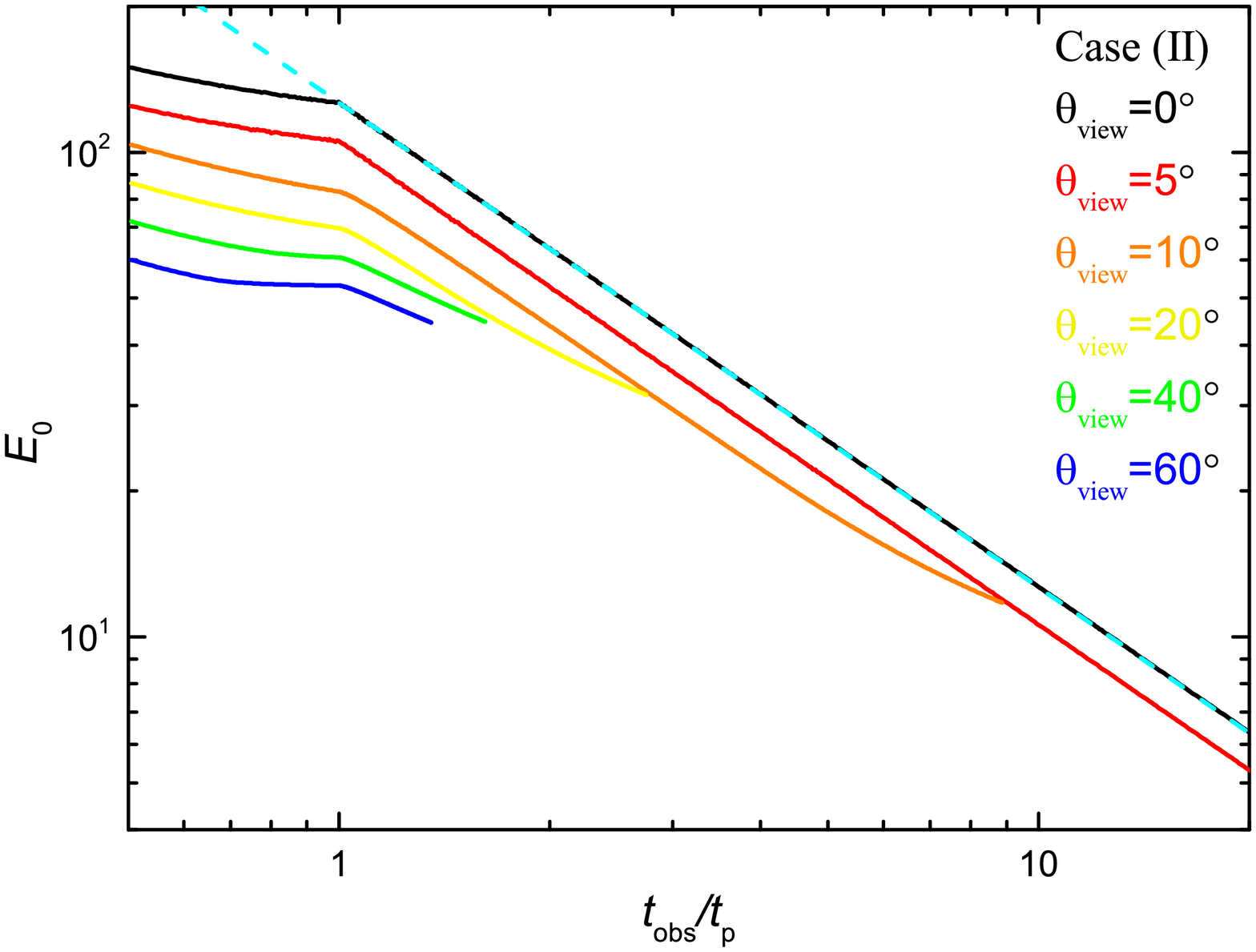} & \includegraphics[angle=0,scale=0.3,trim=00 0 00 0,clip]{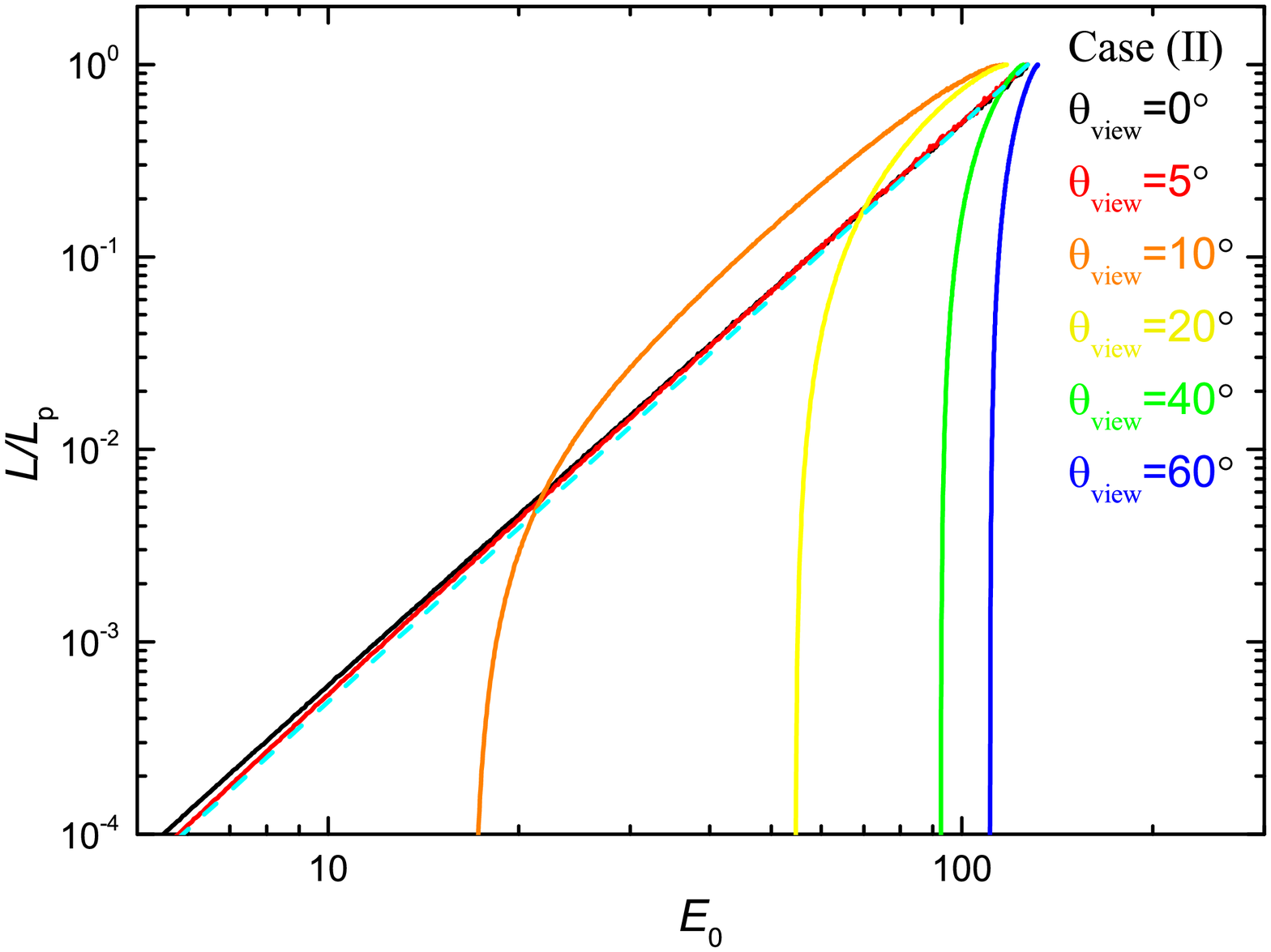}\\
\includegraphics[angle=0,scale=0.3,trim=00 0 00 0,clip]{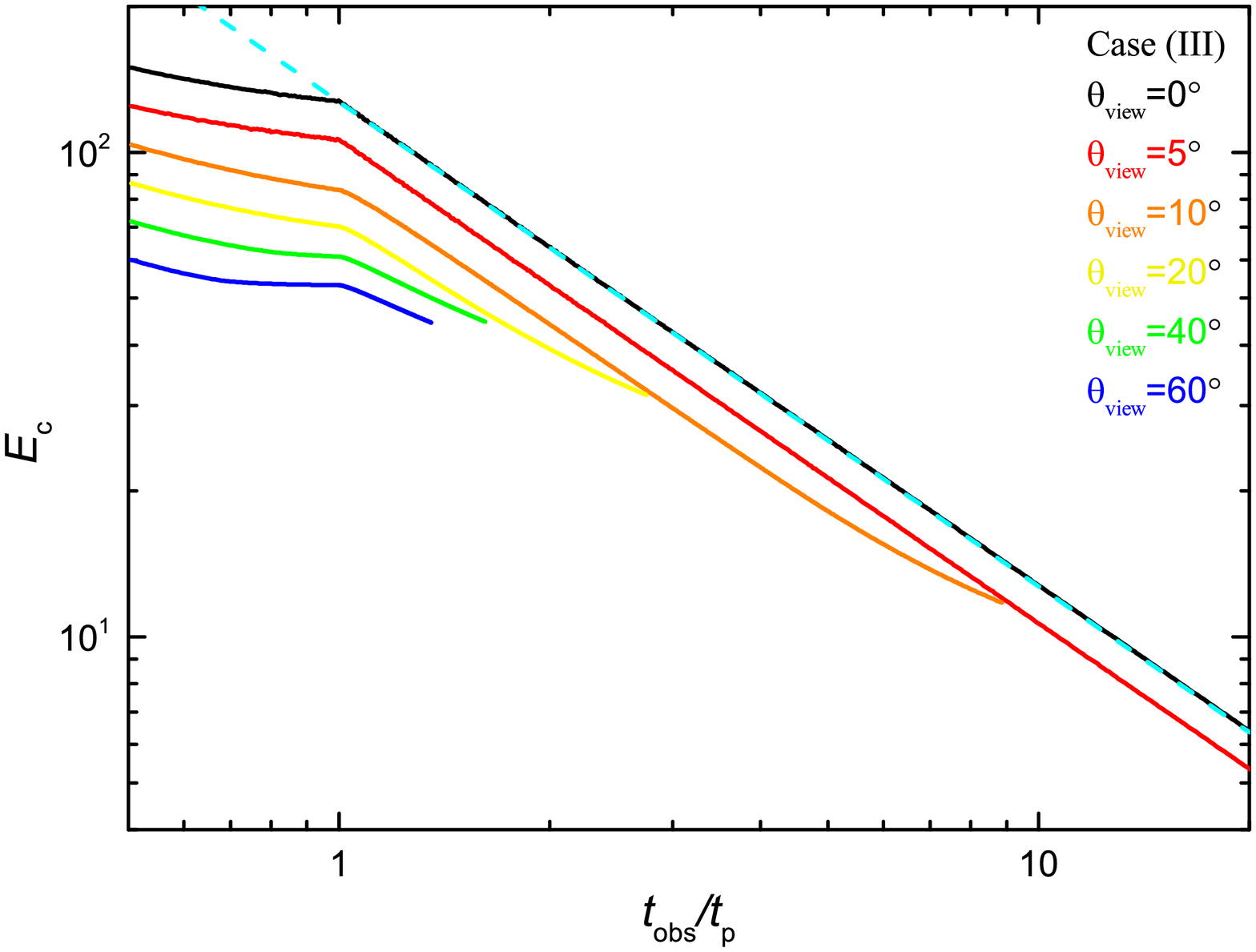} & \includegraphics[angle=0,scale=0.3,trim=00 0 00 0,clip]{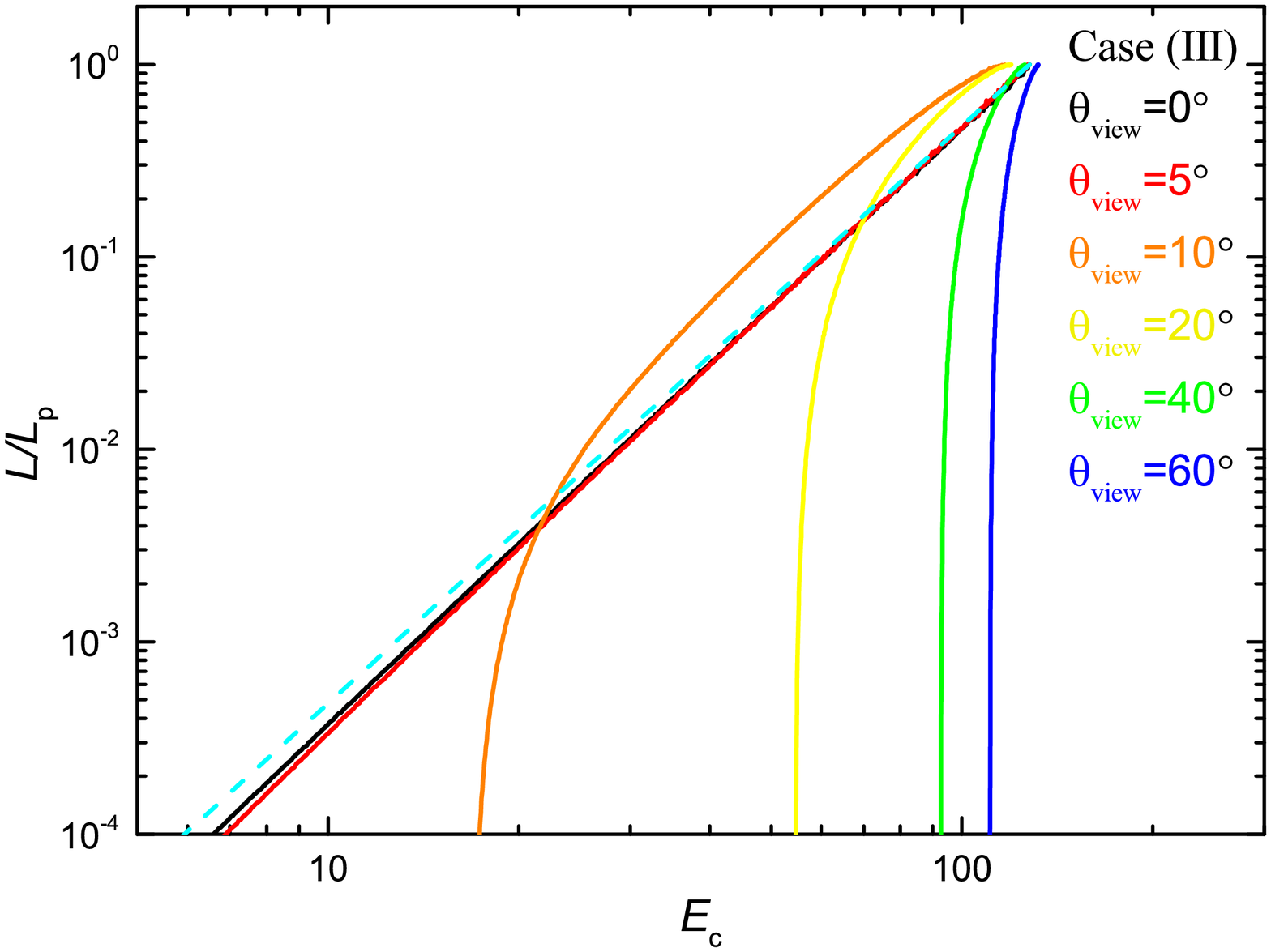}\\
\end{tabular}
\caption{Evolution of $E_0$ ($E_c$) and relations of $L-E_0$ ($L-E_c$).
A 1.2 shift between two adjacent $E_0-t_{\rm obs}$ ($E_c-t_{\rm obs}$) is applied in the plot for clarity.}\label{myfigD}
\end{figure}
%%%%%%%%%%%%%%%%%%%%%%%%%%%%%%%%%%%%%%%%%%%%%%%%%%%%%%%%%%%%00000000000000000000000000000000000000000000000000000000000000000000
%%%%%%%%%%%%%%%%%%%%0%%%%%%%%%%%%%%%%%%%%%%%%%%%%%%%%%%%%%%%%000000000000000000000000000000000000000000000000000000000000000000000

\clearpage
\begin{figure}
\centering
\begin{tabular}{cc}
\includegraphics[angle=0,scale=0.3,trim=00 0 00 0,clip]{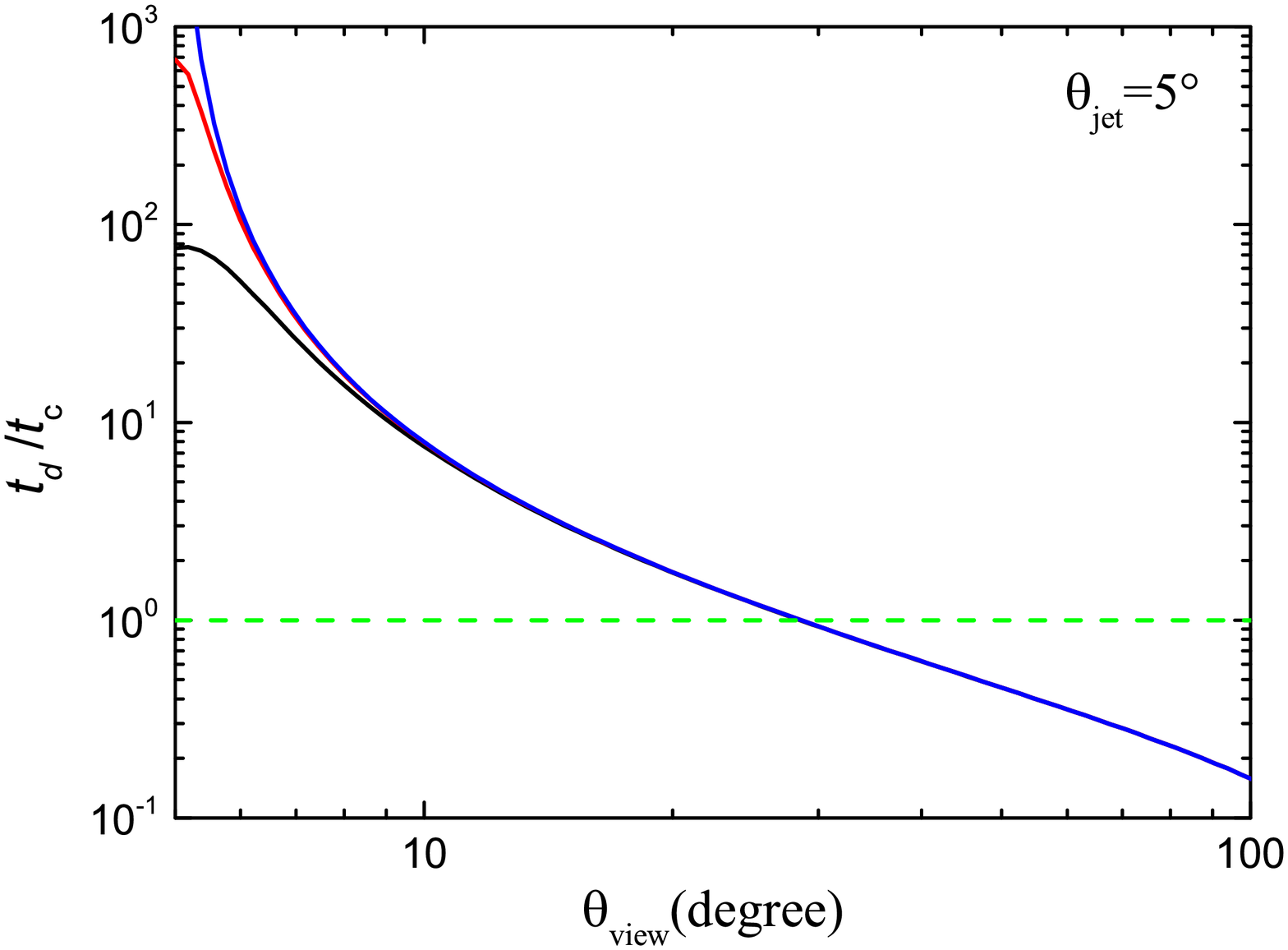} & \includegraphics[angle=0,scale=0.3,trim=00 0 00 0,clip]{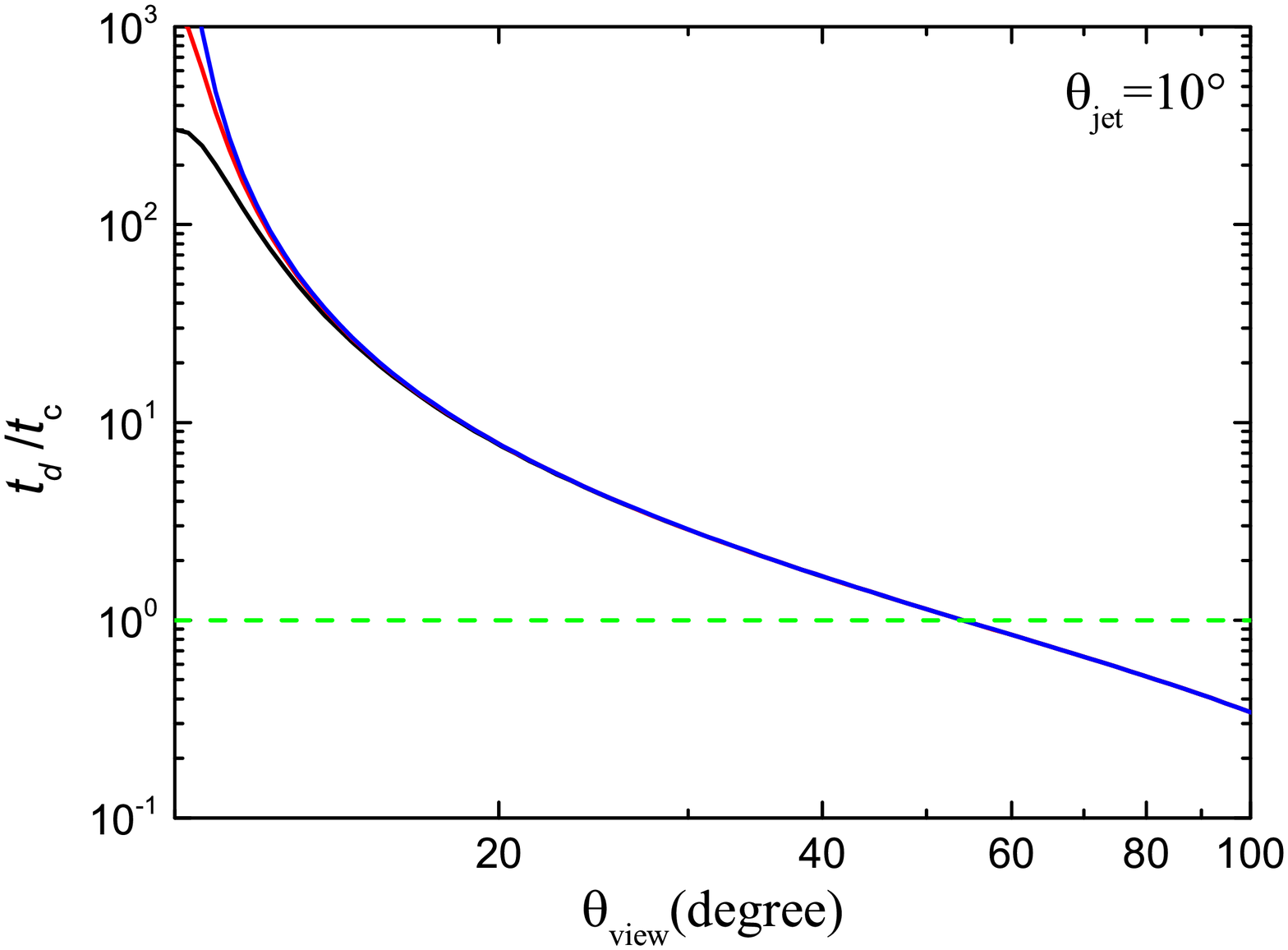}\\
\end{tabular}
\caption{Value of $t_d/t_c$ for different $\theta_{\rm view}$,
where the black, red, and blue solid lines represent the value of $t_d/t_c$
calculated with $\Gamma=50$, 150, 450, respectively.
In addition, $\theta_{\rm jet}=5^\circ$ ($10^\circ$) is adopted in the left (right) panel
and the green dashed lines represent $t_d=t_c$.}\label{myfigE}
\end{figure}
%%%%%%%%%%%%%%%%%%%%%%%%%%%%%%%%%%%%%%%%%%%%%%%%%%%%%%%%%%%%0000000000000000000000000000000000000000000000000000000000000000000000
%%%%%%%%%%%%%%%%%%%%0%%%%%%%%%%%%%%%%%%%0%%%%%%%%%0%%%%%%%%%0%%%000000000000000000000000000000000000000000000000000000000000000000000

\clearpage
\begin{figure}
\centering
\begin{tabular}{c}
\includegraphics[angle=0,scale=0.5,trim=0 0 0 0,clip]{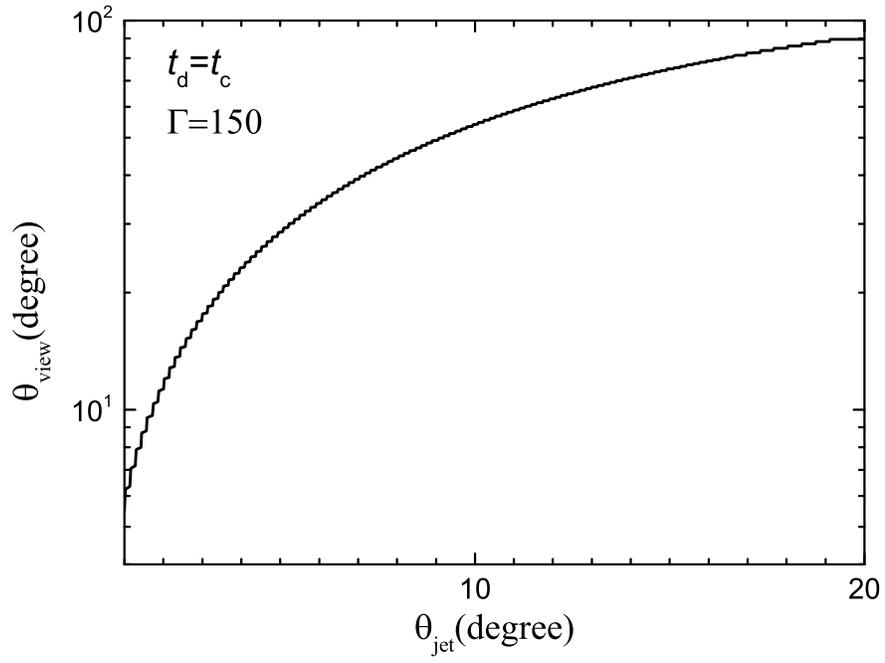}\\
\end{tabular}
\caption{Value of $\theta_{\rm view}$ satisfying $t_d=t_c$ for different $\theta_{\rm jet}$, where $\Gamma=150$ is adopted.}\label{myfigF}
\end{figure}
%%%%%%%%%%%%%%%%%%%%%%%%%%%%%%%%%%%%%%%%%%%%%0%%%%%%%%%%%%%%%000000000000000000000000000000000000000000000000000000000000000000000
%%%%%%%%%%%%%%%%%%%%0%%%%%%%%%%%%%%%%%%%%%%%%%%%%%%%%%%%%%0%%%0000000000000000000000000000000000000000000000000000000000000000000000000

\clearpage
\begin{figure}
\centering
\begin{tabular}{c}
\includegraphics[angle=0,scale=0.5,trim=90 0 90 0,clip]{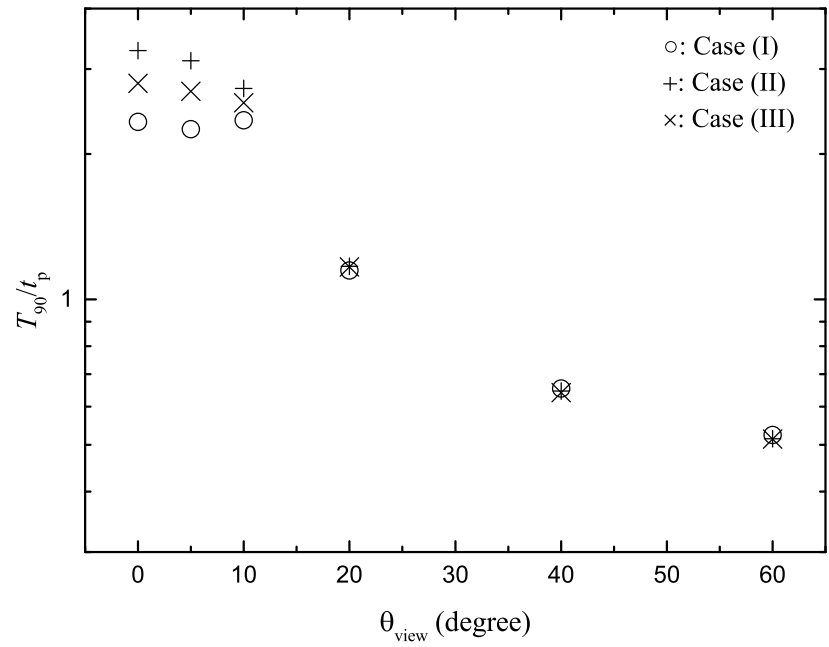}\\
\end{tabular}
\caption{Relation of $T_{90}$ and $t_p$ for different viewing angle,
where the symbols of ``$\circ$'', ``$+$'', and ``$\times$'' represent the results
from the simulations with Case (I), (II), and (III), respectively.}\label{myfigG}
\end{figure}
%%%%%%%%%%%%%%%%%%%%%%%%%%%%%%%%%%%%%%%%%%%%%0%%%%%%%%%%%%%%%000000000000000000000000000000000000000000000000000000000000000000000
%%%%%%%%%%%%%%%%%%%%0%%%%%%%%%%%%%%%%%%%%%%%%%%%%%%%%%%%%%0%%%0000000000000000000000000000000000000000000000000000000000000000000000000
\clearpage
\begin{table}
\begin{center}
\caption{The value of $T_{90}$, $t_p$, and $T_{90}/t_p$ for some identified RS flashes.}\label{MyTabA}
%%%%%%%%%%%%%%%%%%%%%%%%%%%%
\begin{tabular}{c|ccc}
\hline\hline							
GRB	&	$T_{90}$~(s)	&	$t_p$~(s)	&	$T_{90}/t_p$	\\
\hline							
GRB 990123\footnotemark[1]~~	&	165.95	&	44.56	&	3.72 	\\
GRB 041219A\footnotemark[1]	&	2931.68	&	1649.68	&	1.78 	\\
GRB 090102\footnotemark[1]~~	&	270.87	&	60.61	&	4.47 	\\
GRB 090510\footnotemark[2]~~	&	0.24	&	0.29	&	0.83 	\\
GRB 110731A\footnotemark[2]	&	1.52	&	5.49	&	0.28 	\\
GRB 130427A\footnotemark[1]	&	66.25	&	14.22	&	4.66 	\\
GRB 130427A\footnotemark[2]	&	8.99	&	14.39	&	0.62 	\\
GRB 140512A\footnotemark[1]	&	706.65	&	213.39	&	3.31 	\\
\hline							
\end{tabular}
\footnotetext[1]{The value of $T_{90}$ and $t_p$ are estimated with the optical flash.}
\footnotetext[2]{The value of $T_{90}$ and $t_p$ are estimated with the \emph{Fermi}-LAT flash.}
\end{center}
\end{table}

%%%%%%%%%%%%%%%%%%%%%%%%%%%%%%%%%%%%%%%%%%%%%%%%%%%%%%%%%%%%00000000000000000000000000000000000000000000000000000000000000000000
%%%%%%%%%%%%%%%%%%%%0%%%%%%%%%%%%%%%%%%%%%%%%%%%%%%%%%%%%%0%%%000000000000000000000000009000000000000000000000000000000000000000000

\end{document}